\documentclass[aps,prd,twocolumn,showpacs,10pt]{revtex4-1}
\pdfoutput=1
\usepackage{amssymb,amsmath,amsthm}
\usepackage{mathrsfs} 

\usepackage{epsfig}
\usepackage{graphicx}

\DeclareMathOperator{\tr}{tr}

\def\0{{\mathbf{0}}}

\def\mathsf{\tr}

\usepackage{amsfonts}
\usepackage{dsfont}
\usepackage{braket}

\newcommand{\Rset}{\ensuremath{\mathds{R}}}

\newcommand{\Nset}{\ensuremath{\mathds{N}}}

\newcommand{\rmd}{\ensuremath{\mathrm{d}}}
\newcommand{\rme}{\ensuremath{\mathrm{e}}}
\newcommand{\rmi}{\ensuremath{\mathrm{i}}}

\newcommand{\evat}[3][]{\ensuremath{\left.#2\right|^{#1}_{#3}}}

\newcommand{\pderiv}[3][]{\ensuremath{\frac{\partial^{#1} #2}{{\partial #3}^{#1}}}}

\newcommand{\pderivat}[4][]{\ensuremath{\evat{\pderiv[#1]{#2}{#3}}{#4}}}

\newcommand{\hfrac}[2]{\ensuremath{\left.#1\middle/#2\right.}}

\newcommand{\fr}[1]{\ensuremath{\frac{1}{#1}}}

\newcommand{\ahalf}{\ensuremath{\fr{2}}}

\newcommand{\afifth}{\ensuremath{\fr{5}}}
\newcommand{\asixth}{\ensuremath{\fr{6}}}

\newcommand{\bket}[1]{\ensuremath{\left| #1 \right\rangle}}

\newcommand{\proyector}[2]{\ensuremath{\ket{#1} \bra{#2}}}

\newcommand{\sandwich}[3]{\ensuremath{\langle #1 | #2 | #3 \rangle}}

\newcommand{\eref}[1]{(\ref{#1})}

\newcommand{\phih}{\hat \phi}
\newcommand{\Hh}{\hat H}
\newcommand{\mh}{\hat m}
\newcommand{\Uh}{\hat U}
\newcommand{\Ih}{\hat I}
\newcommand{\Ps}{\mathcal P}
\newcommand{\Fs}{\mathcal F}
\newcommand{\Rs}{\mathcal R}
\newcommand{\A}{\mathcal A}
\newcommand{\Hs}{\mathcal H}
\newcommand{\Ns}{\mathcal N}
\newcommand{\Wr}{{\overline W}}
\newcommand{\Wb}{{\mathcal W}}
\newcommand{\dtau}{\Delta \tau}
\newcommand{\xv}{\mathbf{x}}
\newcommand{\wh}{{\hat \omega}}

\begin{document}
\title{Unruh--DeWitt detector event rate for trajectories with time-dependent acceleration}
\author{Luis C.\ Barbado}
\email{luiscb@iaa.es}
\affiliation{Instituto de Astrof\'isica de Andaluc\'ia (CSIC), Glorieta de la Astronom\'ia, 18008 Granada, Spain}
\author{Matt Visser}
\email{matt.visser@msor.vuw.ac.nz}
\affiliation{School of Mathematics, Statistics, and Operations Research, Victoria University of Wellington, PO Box 600, Wellington 6140, New Zealand}

\date{\today}

\begin{abstract}

We analyse the response function of an Unruh--DeWitt detector moving with time-dependent acceleration along a one-dimensional trajectory in Minkowski spacetime. To extract the physics of the process, we propose an adiabatic expansion of this response function. This expansion is also a useful tool for computing the click rate of detectors in general trajectories. The expansion is done in powers of the time derivatives of the acceleration (jerk, snap, and higher derivatives). At the lowest order, we recover a Planckian spectrum with temperature proportional to the acceleration of the detector at each instant of the trajectory. Higher orders in the expansion involve powers of the derivatives of the acceleration, with well-behaved spectral coefficients with different shapes. Finally, we illustrate this analysis in the case of an initially inertial trajectory that acquires a given constant acceleration in a finite time.

\end{abstract}

\pacs{03.70.+k, 04.62.+v, 11.90.+t}

\maketitle

\section{Introduction}

The particle content of the state of a quantum field is an observer-dependent notion. The best known and most extreme example of this dependence is the case of Unruh--DeWitt detectors following constant acceleration trajectories in Minkowski spacetime,  where one or several quantum fields are  set to be in their natural vacuum states. These accelerated detectors, when  coupled to the quantum fields, detect a particle spectrum with thermal Planckian shape and temperature proportional to their acceleration. This is the well-known \emph{Unruh effect,} first proposed by Unruh in~\cite{Unruh:1976db}.

To probe the quantum state of a field, one has to use some form of particle detector. Probably the best known idealized model of a particle detector is the so-called  Unruh--DeWitt detector~\cite{Unruh:1976db,DeWitt:1980hx}. Formally, it consists of a point-like particle with at least two internal energy levels, coupled to a scalar field  by means of a simple monopole (scalar) interaction. One can interpret the probability of the detector  increasing its internal energy by an amount $\hbar \omega$,  due to the interaction with the field, as the probability of the presence of a particle of this same energy $\hbar \omega$ in the quantum field. But the behaviour of a particle detector not only depends on the state of the quantum field, but also on the specific trajectory in spacetime followed by the detector. Thus, this notion of particle will also depend on the trajectory being followed by the detector. For example, if we make the detector follow an eternal constant acceleration trajectory, and set the quantum state for the field to be the vacuum state (vacuum as seen by  inertial observers), we recover the thermal spectrum characteristic of the Unruh effect. 

For trajectories with (eternal) constant acceleration, and in general for (eternal) stationary trajectories, it is clear that the probability of excitation of the detector per unit of time is constant along the trajectory. On the other hand, when we want to deal with more general trajectories, several non-trivial issues arise. The quantity we wish to define is the probability transition rate at some instant of the trajectory. That is, for an ensemble of identical detectors following the same trajectory we wish to calculate the proportion of them that becomes excited per unit time around some instant of the trajectory. 

For a specific implementation of this proposal, one could use for example a switching function controlling the detector interaction, so that the detector is coupled to the field only during some limited period of time. Excitations of the detector could then (loosely) be interpreted as particles detected during this particular period of time. However, in general the introduction of a switching function leads to spurious transitions (even for inertial observers) due to the switching process itself~\cite{Satz:2006kb}. Instead, we will follow a somewhat different procedure, first introduced in~\cite{Schlicht:2003iy}. Using the interaction picture, we formally calculate the probability of transition of a detector coupled to the field (in a constant time-independent manner), but up to some proper time $\tau$. Then, we differentiate with respect to the instant $\tau$ to calculate what we shall call the \emph{response function}, which is the quantity we are looking for. This quantity is  well-defined, and finite, (at least for a wide class of `non-anomalous' trajectories). We will also provide a clear and detailed physical interpretation of this response function in terms of realistic multicomponent detectors, further developing the interpretation pointed out in~\cite{Louko:2006zv}. Nonetheless, the sharp cut-off implicitly generated when evaluating the probability of excitation only up to some specific instant, has the potential to introduce new regularization problems. These problems have been already discussed in the literature~\cite{Schlicht:2003iy,Louko:2006zv,Obadia:2007qf,Satz:2006kb}. We will make use of a formula presented in~\cite{Louko:2006zv} that avoids these problems, and that is particularly useful for our purposes.

A particularly interesting regime for an Unruh--DeWitt detector is that in which it follows a trajectory with a slowly-varying acceleration $g(\tau)$. One expects that in this regime the detector perceives thermal radiation with a slowly-varying temperature, proportional to the acceleration at each instant of the trajectory with the usual Unruh formula $T(\tau)=\hbar |g(\tau)|/(2 \pi c k_B)$~\cite{Unruh:1976db}. We may call it the \emph{adiabatic regime.} If we neglected derivatives of the acceleration higher than the first, the regime would in physical units be defined as the one in which $c \, g'(\tau)/g(\tau)^2 \ll 1$. (This is a constraint on the \emph{jerk} in terms of the instantaneous acceleration.) Setting $c\to1$ as usual, this constraint is commonly rephrased as  $g'(\tau)/g(\tau)^2 \ll 1$~\cite{Kothawala:2009aj,Barcelo:2010pj,Barcelo:2010xk}. As explained in~\cite{Barcelo:2010xk} (in a different and more general context), this is a physically reasonable requirement: If we want the detector to perceive a thermal spectrum with temperature $\propto g(\tau)$, the \emph{relative} change in the acceleration $g'(\tau)/g(\tau)$ during the time needed to detect a particle of the typical energy in the spectrum, $1/g(\tau)$, must be small. In this paper, we will attain this adiabatic regime precisely as a zeroth order approximation of a general asymptotic expansion of the response function of the detector, which we shall call the \emph{adiabatic expansion}. This is an expansion in powers of the derivatives of $g(\tau)$. As we said, a thermal Planckian spectrum with temperature proportional to the acceleration is obtained at lowest order. The other terms will be corrections to that result due to non-adiabaticity, and will become of higher importance the further away we are from the adiabatic regime.

We begin in section~\ref{expressions} with a review of the definitions of the physical quantities involved in the description of an Unruh--DeWitt detector and their physical meaning, followed by a brief discussion of the regularization problems. We also generically describe a one-dimensional trajectory with a time-dependent acceleration. Then, in section~\ref{adiab} we use the expressions introduced in the previous section to calculate the adiabatic expansion of the response function, and in section~\ref{high_energy} we discuss its relation to the high-energy behaviour of the detector. In section~\ref{example}, we use the adiabatic expansion with an example of a time-dependent accelerated trajectory and discuss the results. We conclude in section~\ref{summary} with a brief summary.

\section{General expressions for the transition functions}\label{expressions}

Throughout the article, we will use natural units $\hbar=c=1$. Let us consider a quantized Klein--Gordon real massless scalar field $\phih (x)$, defined in a flat $3+1$ spacetime background, interacting with a localized detector $D$, which carries its own clock measuring proper time $\tau$, and which sweeps out some trajectory $x(\tau)$ through spacetime. The quantum field has an associated Hilbert space $\Hs_F$, and the detector has its own Hilbert space $\Hs_D$. The combined system is defined on the total Hilbert space $\Hs := \Hs_F \otimes \Hs_D$. The quantum field (in the absence of the detector) has Hamiltonian $\Hh_F$, and the detector (when decoupled from the field) evolves according to the Hamiltonian $\Hh_D$.

\subsection{Transition probability}

Apart from the free evolution, the key point is that one postulates an interaction Hamiltonian $\Hh_I$, which couples the detector to the field linearly. That is
\begin{equation}
\Hh_I (\tau) := a \,\mh (\tau) \,\phih\left[x(\tau)\right],
\label{H_inter}
\end{equation}
where $a$ is the coupling constant, and $\mh (\tau)$ is the monopole moment of the detector. For the time being, we do not consider a switching function in~\eref{H_inter}, and thus the detector is switched-on during its entire trajectory.

By assumption, we start in the asymptotic past with the field in its Minkowski vacuum state $\ket{\psi_{F 0}}$, and the detector in its ground state $\ket{E_{D 0}}$ of energy $E_{D 0}$, so that
\begin{equation}
\bket{\psi(\tau \to -\infty)} = \ket{\psi_{F 0}} \otimes \ket{E_{D 0}}.
\label{init_state}
\end{equation}
We will work in the so-called ``interaction picture'' for clarity. The quantum time evolution operator is then given by the time-ordered exponential
\begin{equation}
\Uh (\tau) = \mathscr T \left[ \exp \left( \rmi \int_{-\infty}^\tau \rmd\tau' \Hh_I (\tau') \right) \right].
\label{evol_op}
\end{equation}
If the detector clicks, by jumping to some well defined excited state $\ket{E_D}$, with $E_D > E_{D 0}$, then this forces a collapse
\begin{equation}
\bket{\psi(\tau)} \to \ket{\psi_F} \otimes \ket{E_D}.
\label{collapse}
\end{equation}
The amplitude for this to happen is
\begin{equation}
\A_{\psi_F,E_D}(\tau) = \rmi \bra{\psi_F} \otimes \bra{E_D} \Uh (\tau) \ket{\psi_{F 0}} \otimes \ket{E_{D 0}},
\end{equation}
and the probability that the detector is found in its state $|E_D\rangle$ after an internal time $\tau$ has passed, given that you are \emph{not looking} at the final state of the field, is
\begin{equation}
\Ps_{E_D} (\tau) := \sum_i \left| \A_{\psi_{Fi},E_D}(\tau) \right|^2,
\label{prob_sum}
\end{equation}
where the sum runs over a basis of the Hilbert space $\Hs_F$ of the field.

This probability, by definition, is a non-negative quantity smaller or equal than unity. During its evolution in $\tau$, it can do nothing but oscillate. However, one would like to have a quantity capturing the physical intuition that, during periods in which an acceleration is present, the detector would be continuously encountering new particles, producing some sort of  cumulative effect. 

\subsection{Cumulative macro-detector}

Let us consider $N$ identical Unruh--DeWitt detectors, all of them prepared in the same initial state (in the manner described above) and following identical trajectories. Physically, one might construct this by spreading the $N$ micro-detectors into a tiny volume, so that the largest distance between two detectors is still much smaller that the minimum wavelength one wishes to measure. In this way, one can still consider the whole ensemble as pointlike. This would be a physical model for a more realistic macro-detector. Now, assume in addition that each micro-detector has a coupling constant $\epsilon := a/ \sqrt{N}$. Given a specific $\tau$, one can always take $N$ large enough so that one can approximate the evolution operator by its first-order term in a perturbative expansion in the small constant $\epsilon \ll 1$. In this case we have
the evolution for each individual component of the macro-detector
\begin{multline}
\ket{\psi(\tau)} = \Uh (\tau) \ket{\psi_{F 0}} \otimes \ket{E_{D 0}} \simeq \\
\left[ \Ih + \rmi \int_{-\infty}^\tau \rmd\tau' \Hh_I (\tau') \right] \ket{\psi_{F 0}} \otimes \ket{E_{D 0}},
\label{evolution}
\end{multline}
where $\Ih$ is the identity operator, so that the transition amplitude can be written as
\begin{multline}
\A_{\psi_F,E_D}(\tau) = \\
\rmi \bra{\psi_F} \otimes \bra{E_D} \left[ \int_{-\infty}^\tau \rmd\tau' \Hh_I (\tau') \right] \ket{\psi_{F 0}} \otimes \ket{E_{D 0}} = \\
\rmi \epsilon \sandwich{E_D}{\mh (0)}{E_{D 0}} \int_{-\infty}^\tau \rmd\tau' \rme^{\rmi (E_D-E_{D_0}) \tau'} \\
\times \sandwich{\psi_F}{\phih\left[x(\tau')\right]}{\psi_{F 0}}.
\label{amplitude3}
\end{multline}
Here we have used equation~\eref{H_inter} and the fact that the detector, left to its own devices, evolves according to
\begin{equation}
\mh (\tau) = \rme^{\rmi \Hh_D \tau}\, \mh(0)\, \rme^{-\rmi \Hh_D \tau}.
\label{detector_evol}
\end{equation}
Then, taking into account that 
\begin{equation}
\sum_{i} \proyector{\psi_{Fi}}{\psi_{Fi}} = \Ih_{\Hs_F},
\label{identity}
\end{equation}
where $\Ih_{\Hs_F}$ is the identity operator in the Hilbert space $\Hs_F$, the transition probability can be written as
\begin{multline}
\Ps_{E_D} (\tau)
= \epsilon^2 \left| \sandwich{E_D}{\mh (0)}{E_{D 0}} \right|^2 \\
\times \int_{-\infty}^\tau \rmd\tau'' \int_{-\infty}^\tau \rmd\tau' \;\rme^{-\rmi (E_D-E_{D_0}) (\tau'' - \tau')} \;\Wb (\tau'',\tau').
\label{prob2}
\end{multline}
Here, $\Wb(\tau'',\tau')$ is the Wightman distribution written in terms of the proper time of the trajectory,
\begin{equation}
\Wb (\tau'',\tau') := \sandwich{\psi_{F 0}}{\phih(x(\tau'')) \; \phih(x(\tau'))}{\psi_{F 0}},
\label{w-tau}
\end{equation}
which can be explicitly expressed as~\cite{birrell1984quantum}
\begin{multline}
\Wb (\tau'',\tau') = \\
-\fr{4 \pi^2 \left\{ \left[ t(\tau'') - t(\tau') - \rmi \,\varepsilon \right]^2 - \left[ \xv(\tau'') - \xv(\tau') \right]^2 \right\}},
\label{w-expl}
\end{multline}
where the limit $\varepsilon \to 0^+$ is implicitly taken \emph{after integrating} the function.

As a final step let us define the quantity $\Ns_{E_D}(\tau) := N \Ps_{E_D}(\tau)$. Then
\begin{multline}
\Ns_{E_D}(\tau) = a^2 \left| \sandwich{E_D}{\mh (0)}{E_{D 0}} \right|^2 \\
\times \int_{-\infty}^\tau \rmd\tau'' \int_{-\infty}^\tau \rmd\tau'\;  \rme^{-\rmi (E_D-E_{D_0}) (\tau'' - \tau')} \; \Wb (\tau'',\tau').
\label{number-detectors}
\end{multline}
This can be interpreted as the average number of micro-detectors within the macroscopic detector that are excited at time $\tau$. This quantity is no longer conceptually restricted to have a value smaller than one and in fact does exhibit cumulative effects.

In the situations analysed in this paper the trajectories are always inertial in the past but are extended up to arbitrarily long time intervals. The previous formula can always be used under the assumption that in the limit $\Delta T \to \infty$ (where $\Delta T$ is the observation time) the number of finally excited micro-detectors $\Ns_{E_D}$ (which in general will grow with $\Delta T$, possibly even going also to infinity), divided by the total number of micro-detectors $N$, is such that $\Ns_{E_D} / N \ll 1$.

\subsection{Response function (and click rate)}

The quantity we have calculated so far is the average number of detectors being in an excited state of energy $E_D$ at some time $\tau$. But the quantity that we are really interested in is the one that gives information about the amount of particles that ``appear to the eyes'' of an observer following a specific trajectory at a certain time, i.e. independently of the details of the detector.
In this sense, we should note that the average number is the product of two factors with entirely different origins. The first one depends only on the characteristics of the detector. The second is the quantity~\cite{birrell1984quantum}
\begin{equation}
\Fs (\omega, \tau) := \int_{-\infty}^\tau \rmd\tau'' \int_{-\infty}^\tau \rmd\tau' \;\rme^{-\rmi \omega (\tau'' - \tau')} \;\Wb (\tau'',\tau'),
\label{response-function}
\end{equation}
with $\omega := E_D - E_{D_0}$. This quantity is ``detector independent'', and retains the cumulative property of the average number. (This function $\Fs$ evaluated at $\tau \to \infty$ is called the \emph{response function} in Birrell \& Davies~\cite{birrell1984quantum}, but we will reserve this name for its time derivative, as done in other sources~\cite{Satz:2006kb,Louko:2006zv}.) To better understand the behaviour of the detector at a certain moment of time one can calculate instead the time derivative of this quantity:
\begin{equation}
\Rs (\omega, \tau) := \pderiv{\Fs}{\tau} (\omega, \tau).
\label{rate-def}
\end{equation}
Explicitly, this can be easily calculated to be
\begin{equation}
\Rs (\omega, \tau) = 2~\Re \left[ \int_{-\infty}^0 \rmd s\; \rme^{-\rmi \omega s} \; \Wb(\tau+s,\tau) \right].
\label{rate-expl}
\end{equation}
This is what we  shall call the \emph{response function} of an Unruh--DeWitt detector. As we can see, the quantity only depends on the past history of the detector, and thus is completely causal~\cite{Schlicht:2003iy}.

In the case of eternally stationary trajectories, either inertial or constant acceleration, this can be further simplified. We have then that the Wightman distribution is invariant under time translations along the trajectory, so that one can define
\begin{multline}
\Wb(s) := \Wb (s,0) = \Wb (\tau+s,\tau) = \\
\Wb(\tau+\Delta \tau + s,\tau + \Delta \tau).
\end{multline}
Furthermore
\begin{equation}
\Wb^*(s) = \Wb^*(s,0) = \Wb (0,s) = \Wb (-s,0) = \Wb(-s).
\end{equation}
This allows us to obtain the following particularly simple expression for this case:
\begin{align}
\Rs (\omega) = \int_{-\infty}^\infty \rmd s\; \rme^{-\rmi \omega s} \;\Wb(s).
\label{rate-expl-invariant}
\end{align}
This expression fits the usual one for the response function dealing with a stationary situation, as is the case of a constant acceleration trajectory~\cite{birrell1984quantum}. For this reason, one could be tempted to call this function \emph{click-rate function}, characterizing the numbers of excitations produced per unit of time. However, this interpretation is not always completely appropriate as one can check that this function can sometimes attain negative values, indicating de-excitation of the detector, typically in moments of decreasing acceleration. Nevertheless, as we will show below, its behaviour provides important insights regarding the time development of the detection process, and therefore in the particle perception process (see also the discussion in~\cite{Louko:2006zv}). One could attempt to build more complicated models of more realistic detectors by incorporating notions of irreversibility and detector latency, but such refinements do not seem crucial to the questions we will explore.

\subsection{Finite-time detectors and the regularized Wightman function}\label{real_detectors}

In this paper, we are considering a multi-part macro-detector that is always switched on, and we measure average number of micro-detectors excited at some time $\tau$. If we instead wanted to consider a detector with a different switching time or process, we should have written
\begin{equation}
\Hh'_I (\tau') := a~\xi(\tau')\; \mh (\tau') \;\phih\left[x(\tau')\right],
\label{H_inter_sw}
\end{equation}
instead of~\eref{H_inter} for the interaction term in the Hamiltonian, where  $\xi(\tau')$ is now a \emph{switching function,} that takes positive values during the interaction and vanishes during the periods of no interaction (see for example~\cite{sriramkumar:1996} for a description of Unruh--DeWitt detectors with different switching functions). This function should be integrated together with the Wightman distribution when calculating the response function.

As pointed out in~\cite{Satz:2006kb}, the Wightman distribution is strictly speaking well defined only when integrated with a \emph{bump function,} that is, a smooth function of compact support. In fact, this criterion is reasonable, as a physically realizable detector is never switched-on during an infinite amount of time, and it is never switched on or off in a (strictly) sharp manner either. We will call \emph{finite-time detectors} the detectors with a bump function as the switching function.

One can see that calculating the response function by using~\eref{rate-expl} corresponds to choosing the switching function
\begin{equation}
\xi(\tau') = \Theta (\tau-\tau'),
\label{switching}
\end{equation}
where $\Theta$ is the Heaviside step function, i.e. the detector is on from the infinite past until a proper time $\tau$. This function does not fulfil any of the two requirements of a bump function. Thus, results obtained using~\eref{rate-expl} together with~\eref{w-expl} are not strictly speaking correct. For example, Schlicht~\cite{Schlicht:2003iy} showed that this procedure leads to non-Lorentz invariant results, giving non-zero detection rates even for inertial observers.

However, in~\cite{Satz:2006kb} it is also proven that we can approach with arbitrary precision the regime with an infinite switching time, and a sharp switch-off process, by using a concrete family of bump functions as switching functions. Consider a finite-time detector that starts switched-off in the asymptotic past [$\xi(\tau') = 0$], then it is turned on smoothly during a period $\delta$ up to some time $\tau_0$, remains switched on [$\xi(\tau') = 1$] during a period $\Delta T$ up to some time $\tau$, and is finally turned off again smoothly during a period $\delta$. One can calculate the quantity $\Ns_{E_D}$ in this process, and with it the response function by differentiating with respect to the switching-off time $\tau$. This last quantity can be approximated by~\cite{Satz:2006kb}
\begin{multline}
\Rs_{\rm finite-time} (\omega, \tau) = -\frac{\omega}{4\pi} \\
-\fr{2 \pi^2} \int_{-\Delta T}^0 \rmd s \left\{ \frac{\cos (\omega s)}{ \left[ t(\tau+s) - t(\tau) \right]^2 - \left[ \xv(\tau+s) - \xv(\tau) \right]^2 } \right. \\
\left. - \fr{s^2} \right\} +\fr{2\pi^2 \Delta T} + O\left(\frac{\delta}{\Delta T^2}\right).
\label{real_detector}
\end{multline}
\newpage
In the limit where $\Delta T \to \infty$ and $\delta \to 0$, equation \eref{real_detector} approaches the following value:
\begin{equation}
\Rs (\omega, \tau) = 2 \int_{-\infty}^0 \rmd s\; \cos (\omega s) \; W (\tau+s,\tau).
\label{rate-reg}
\end{equation}
This is identical to equation~\eref{rate-expl}, but replacing the Wightman distribution $\Wb$ by what we shall call the \emph{regularized Wightman function,} which is defined by~\cite{Satz:2006kb}
\begin{multline}
W (\tau'',\tau') := -\fr{4 \pi^2} \\
\times \left\{ \fr{ \left[ t(\tau'') - t(\tau') \right]^2 - \left[ \xv(\tau'') - \xv(\tau') \right]^2 } - \fr{(\tau''-\tau')^2} \right\}.
\label{W-reg}
\end{multline}
This regularized quantity is well-behaved in the limit where the two arguments coincide $\tau''\to\tau'$. 
If we compare~\eref{real_detector} with~\eref{rate-reg}, we explicitly obtain
\begin{multline}
\left| \Rs_{\rm finite-time} (\omega, \tau) - \Rs (\omega, \tau) \right| = \\
\left| \fr{2 \pi^2} \int_{-\infty}^{-\Delta T} \rmd s \left\{ \frac{\cos (\omega s)}{ \left[ t(\tau+s) - t(\tau) \right]^2 - \left[ \xv(\tau+s) - \xv(\tau) \right]^2 } \right. \right. \\
\left. \left. - \fr{s^2} \right\} +\fr{2\pi^2 \Delta T} + O\left(\frac{\delta}{\Delta T^2}\right) \right| \\
\leq \frac{3}{2 \pi^2 \Delta T} + O\left(\frac{\delta}{\Delta T^2}\right).
\label{compare_rates}
\end{multline}
This result means that, insofar as the detector is switched on during a sufficiently long period of time ($\Delta T \to \infty$), and the switching timescale is sufficiently short ($\delta \to 0$), one can reach arbitrary precision using equation~\eref{rate-reg} for calculating the response function of this finite-time detector. Our calculations with~\eref{rate-reg} will then have this precise physical interpretation.

Also, note that the quantity~\eref{W-reg} is not merely a distribution but also well-defined as a \emph{function}. It needs no $\varepsilon \to 0^+$ limit calculation, as the pole at $\tau''=\tau'$ is `dodged' by subtracting the function $1/(\tau''-\tau')^2$. It is thereby analytically much more tractable, and thus much more useful for our purposes. Note also that it is manifestly Lorentz invariant.

\subsection{Variable acceleration trajectories}

We are particularly interested in calculating the response function of a one-dimensional trajectory with time-dependent acceleration. For this class of trajectories, we can explicitly write the integrated equations for the coordinates (see for example~\cite{Padmanabhan:2003gd,moller_theory_1976}). If we consider for simplicity a trajectory along the $x$-axis, we have:
\begin{align}
t(\tau) &= t_0 + \int_{\tau_0}^\tau \rmd \tau' \cosh \chi (\tau'), \nonumber\\
x(\tau) &= x_0 + \int_{\tau_0}^\tau \rmd \tau' \sinh \chi (\tau'), \nonumber\\
y &= y_0, \quad z = z_0.
\label{traj-acc}
\end{align}
Here
\begin{equation}
\chi (\tau) := \chi_0 +  \int_{\tau_0}^\tau \rmd \tau' g(\tau').
\label{xi-def}
\end{equation}
The function $g(\tau)$ is the instantaneous proper acceleration of the trajectory along the $x$-axis, that is, as measured in an inertial and comoving reference frame. $(t_0,x_0,y_0,z_0)$ are the initial coordinates at $\tau=\tau_0$. Finally the initial velocity at $\tau_0$ is $(\cosh \chi_0, \sinh \chi_0, 0, 0)$. Trajectories described in this way are also used in~\cite{Kothawala:2009aj}, where they also study the response of Unruh--DeWitt detector with time-dependent acceleration. However, their choice of the ``time variable'' with respect to which the response function is defined differs from ours and, actually, leads to a non-causal dependence of the transition rate: The detector, after having stayed in an inertial trajectory during its whole past history, could eventually click just because ``it is prepared'' to start to accelerate some time \emph{after} the click. Because of this non-causal aspect to their calculation, their results are not directly comparable to ours.

Inserting~\eref{traj-acc} and~\eref{xi-def} in equation~\eref{W-reg}, we obtain the expression
\begin{multline}
W(\tau+s,\tau;g] = 
\fr{4 \pi^2 s^2} \\
- \fr{4 \pi^2} \left\{ \left[ \int_0^s \rmd s' \cosh \left( \int_0^{s'} \rmd s'' g(\tau + s'') \right) \right]^2 \right. \\
- \left. \left[ \int_0^s \rmd s' \sinh \left( \int_0^{s'} \rmd s'' g(\tau + s'') \right) \right]^2 \right\}^{-1}.
\label{Wreg-g}
\end{multline}
Here we have written the functional dependence on the acceleration $g$ explicitly in the argument of the regularized Wightman function.  Specifically, observe that the notation $W(\_\_,\_\_;\_\_]$ is a completely standard (but somewhat rare) notation designed to emphasise that this object is a \emph{function} of the first two arguments, but a \emph{functional} of the third argument. 
Note that none of the initial conditions of the trajectory are present in this expression. This is completely natural, as they merely correspond to the choice of the 4-velocity and origin of an inertial reference frame, and thus play no role in calculating the response function, which is Poincar\'e  invariant.

\section{Adiabatic expansion of the response function}\label{adiab}

Equations~\eref{rate-reg} and~\eref{Wreg-g} formally are enough to completely calculate the response function for an arbitrary acceleration history $g(\tau)$. But except for very simple acceleration histories, because of the integrals involved in this expressions, one will not be able to calculate explicitly even the regularized Wightman function in~\eref{Wreg-g}. However, as announced in the introduction, we can use now equations~\eref{rate-reg} and~\eref{Wreg-g} to calculate the adiabatic expansion of the response function in~\eref{rate-reg}. This expansion will provide an arbitrarily good approximation to the exact value of the response function in the limit in which the acceleration varies slowly enough, in a sense to be precisely defined below. Note that, in order to calculate the numerical value of $\Rs (\omega, \tau)$, we have to know the whole past history of the detector, via the acceleration history $g(\tau')$. As we will see, within the framework of an adiabatic expansion we can appropriately calculate the response function just using local properties of the trajectory at some time $\tau$. With this aim, let us assume that the acceleration is a real analytic function $g(\tau')$, at least for $\tau' \le \tau$. In order to obtain an asymptotic series expansion of the response function $\Rs$, let us introduce an auxiliary parameter $\alpha$, and define a new auxiliary function $\Wr$ in the following way: Substitute 
\begin{equation}
g(\tau') \to g\left(\alpha(\tau'-\tau) + \tau\right),
\label{new-g}
\end{equation}
in the formal expression of the Wightman function~\eref{Wreg-g} so that
\begin{multline}
\Wr \left(\tau+s,\tau,\alpha;g\right] := \fr{4 \pi^2 s^2} \\
- \fr{4 \pi^2} \left\{ \left[ \int_0^s \rmd s' \cosh \left( \int_0^{s'} \rmd s'' g(\tau + \alpha s'') \right) \right]^2 \right. \\
\left. - \left[ \int_0^s \rmd s' \sinh \left( \int_0^{s'} \rmd s'' g(\tau + \alpha s'') \right) \right]^2 \right\}^{-1}.
\label{new-Wreg}
\end{multline}
It is obvious that the regularized Wightman function $W$ given in equation~\eref{Wreg-g} is obtained from this auxiliary function by setting $\alpha=1$.

We can find a \emph{formal} series expansion  for~\eref{new-Wreg} in powers of $\alpha$. The coefficients of this series expansion will depend on $s$ and on increasingly higher derivatives of $g$ evaluated at $\tau$. Explicitly
\begin{multline}
\Wr \left(\tau+s,\tau,\alpha;g\right] = W_0 \left(s,g(\tau)\right) + W_1 \left(s,g(\tau),g'(\tau)\right) \alpha \\
+ W_2 \left(s,g(\tau),g'(\tau),g''(\tau)\right) \alpha^2 + \ldots,
\label{W-series}
\end{multline}
where
\begin{equation}
W_0 \left(s,g(\tau)\right) := \lim_{\alpha \to 0^+} \Wr \left(\tau+s,\tau,\alpha;g\right],
\end{equation}
and
\begin{multline}
W_n \left(s, g(\tau),g'(\tau),\ldots,g^{(n)}(\tau) \right) := \\
\fr{n!} \pderivat[n]{\Wr \left(\tau+s,\tau,\alpha;g\right]}{\alpha}{\alpha \to 0^+}.
\label{Wn_terms_def}
\end{multline}
Once we have performed the expansion in powers of $\alpha$, we \emph{formally} evaluate the expression in $\alpha = 1$ to recover the value of the Wightman function in~\eref{Wreg-g}, (this is straightforward from~\eref{new-Wreg}), but now written as an asymptotic series expansion: 
\begin{multline}
W (\tau+s,\tau;g] \sim W_0 \left(s,g(\tau)\right) + W_1 \left(s,g(\tau),g'(\tau)\right) \\
+ W_2 \left(s,g(\tau),g'(\tau),g''(\tau)\right) + \ldots.
\label{W-series-final}
\end{multline}
Lastly, we will (again \emph{formally}) calculate the integral in~\eref{rate-reg} term by term using~\eref{W-series-final}. It is clear that each term in the series is integrable: First, for each value of $s$, the object $\Wr (\tau+s,\tau,\alpha;g]$ is a real analytic function of $\alpha$, and thus the coefficients $W_n$ are finite for each $s$. Second all the coefficients tend to zero for $s \to -\infty$ faster than $1/s^2$ because, for $\alpha$ small enough in~\eref{W-series}, taking just a finite number of terms in the series in $\alpha$ should provide an arbitrarily good approximation to $W (\tau+s,\tau;g]$, which actually vanishes in that limit faster than $1/s^2$, as can be easily checked. We then have
\begin{multline}
\Rs (\omega, \tau) \sim \Rs_0 \left(\omega,g(\tau)\right) + \Rs_1 \left(\omega,g(\tau),g'(\tau)\right) \\
+ \Rs_2 \left(\omega,g(\tau),g'(\tau),g''(\tau)\right) + \ldots.
\label{rate-series}
\end{multline}
where
\begin{multline}
\Rs_n \left(\omega, g(\tau),g'(\tau),\ldots,g^{(n)}(\tau) \right) := \\
2 \int_{-\infty}^0 \rmd s~\cos(\omega s) \; W_n \left(s, g(\tau),g'(\tau),\ldots,g^{(n)}(\tau) \right).
\label{rate-coef}
\end{multline}
Expression~\eref{rate-series} is what we shall call the \emph{adiabatic expansion of the response function.}

Let us remark that expression~\eref{rate-series} is not merely a short time expansion of $\Rs$ around $\tau$ and, although closely related (as we will see later), it is also conceptually and mathematically different from a high-energy expansion in inverse powers of $\omega$.
In particular, the first four coefficients of the adiabatic expansion are
\begin{widetext}
\begin{align}
\Rs_0 \left(\omega,g(\tau)\right) = & |g(\tau)| \frac{\wh}{2 \pi} \; \fr{\rme^{2 \pi \wh} - 1}, \nonumber \\[10pt]
\Rs_1 \left(\omega,g(\tau),g'(\tau)\right) = & g(\tau) \left[ \frac{g'(\tau)}{g(\tau)^2} \; f_{1,1} (\wh) \right], \nonumber \\[10pt]
\Rs_2 \left(\omega,g(\tau),g'(\tau),g''(\tau)\right) = & |g(\tau)| \left[ \frac{g''(\tau)}{g(\tau)^3} \; f_{2,1} (\wh) + \frac{g'(\tau)^2}{g(\tau)^4} \; f_{2,2} (\wh) \right], \nonumber \\[10pt]
\Rs_3 \left(\omega,g(\tau),g'(\tau),g''(\tau),g'''(\tau)\right) = & g(\tau) \left[ \frac{g'''(\tau)}{g(\tau)^4}\; f_{3,1} (\wh) + \frac{g'(\tau) g''(\tau)}{g(\tau)^5} \; f_{3,2} (\wh) + \frac{g'(\tau)^3}{g(\tau)^6} \; f_{3,3} (\wh) \right].
\label{series-terms}
\end{align}
\end{widetext}
Here $\wh := \omega/|g(\tau)|$, and the $f_{i,j}$ are dimensionless functions of the dimensionless variable $\wh$. The first term is simply the thermal spectrum for a temperature $T = \hfrac{\left|g(\tau)\right|}{2 \pi k_B}$. That is, as we already said, we recover a thermal Planckian spectrum detection as a zeroth-order approximation in the adiabatic expansion. Note that $\Rs_0$ depends only on the acceleration, while $\Rs_1$ also depends on the \emph{jerk,} and $\Rs_2$ also depends on the \emph{snap,} etcetera. Moreover, $\Rs_0$ actually depends only on the \emph{absolute value} of the acceleration. This is what one should expect, as the sign of the acceleration $g(\tau)$ \emph{by itself} has no physical meaning in Minkowski spacetime. On the other hand, the term $\Rs_1$ depends on the sign of the quotient of the jerk by the acceleration, $g'(\tau)/g(\tau)$. But this sign does have a physical meaning, indicating an increasing or decreasing acceleration. That is the reason for the $\Rs_0$ term to be multiplied by $|g(\tau)|$, while the $\Rs_1$ term is multiplied by $g(\tau)$. This sign dependency goes alternatively on and off with odd and even terms in the expansion, respectively, in the expressions~\eref{series-terms}. Apart from this alternating factor, the quantities between brackets in~\eref{series-terms} are sums of dimensionless functions $f_{i,j}$ multiplied by powers of the dimensionless quantities $\{ g^{(l)}(\tau)/g(\tau)^{l+1} \}$.

The explicit expressions for the first three $f_{i,j} (\wh)$ functions are
\begin{align}
f_{1,1} (\wh) = & -\fr{4 \pi^2} \left[1 + 2 \wh~F(\wh) + \wh^2~G(\wh) \right] ; \nonumber \\[10pt]
f_{2,1} (\wh) = & -\frac{\wh}{24 \sinh^4 (\pi \wh)} \left\{ \wh \left[ \left(3+4\pi^2\right) \right. \right. \nonumber \\
& \left. \left. + \left(-3+2\pi^2\right) \cosh(2 \pi \wh) \right] - 3 \pi \sinh (2 \pi \wh) \right\}; \nonumber \\[10pt]
f_{2,2} (\wh) = & \fr{12 \sinh^2(\pi \wh)} \left\{ -1 +3 \wh^2\left(-3+4 \pi^2\right) \right. \nonumber \\
& \left. + \pi \wh \left[\left(-8+\wh^2\left(1-3\pi^2\right)\right)\coth(\pi \wh) \right. \right. \nonumber \\
& \left. \left. - 9 \pi \wh \left(-2 + \pi \wh \coth(\pi \wh) \right) \sinh^{-2}(\pi \wh) \right] \right\},
\label{f-ij}
\end{align}
where we have defined the functions
\begin{align}
F(\wh) := & \int_{-\infty}^0 \rmd u \frac{u \sin (\wh u)}{1-\rme^{-u}}; \nonumber\\
G(\wh) := & \int_{-\infty}^0 \rmd u \frac{u^2 \cos (\wh u)}{1-\rme^{-u}}.
\label{funcs_F_G}
\end{align}
Explicit evaluation of these integrals leads to formulae in terms of the digamma function $\psi(x):=\Gamma'(x)/\Gamma(x)$:
\begin{align}
F(\wh) = & \frac{\rmi}{2} \left[ \psi'(1-\rmi \wh) - \psi'(1+\rmi \wh) \right];  \nonumber\\\
G(\wh) = & \ahalf \left[ \psi''(1-\rmi \wh) + \psi''(1+\rmi \wh) \right].
\label{digamma}
\end{align}

\begin{figure}[ht]
\centering
\includegraphics[width=0.9\linewidth]{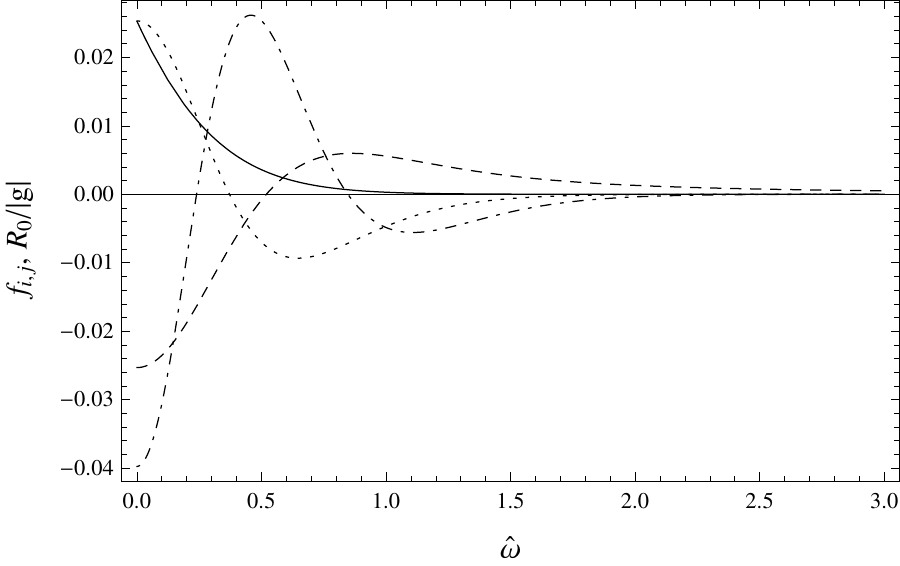}
\caption{Values of $\Rs_0 \left(\omega,g(\tau)\right) / |g(\tau)|$ (the normalized thermal spectrum, solid line), $f_{1,1} (\wh)$ (dashed line), $f_{2,1} (\wh)$ (dotted line) and $f_{2,2} (\wh)$ (dash-dot line), as functions of $\wh$.}
\label{spectrums}
\end{figure}

In figure~\ref{spectrums}, we plot the values of the first three $f_{i,j} (\wh)$ functions, together with the (normalised) thermal spectrum $\Rs_0 \left(\omega,g(\tau)\right) / |g(\tau)|$. Note that each term contributing to the adiabatic series has a well-defined shape in terms of the quantity $\wh$. Changing the value of the derivatives of $g(\tau)$ only increases or decreases the overall contribution of each term (but not its shape) in the total output. That is, we can consider the total response function as a superimposed sum of different spectra. The first one is the thermal spectrum. The others have different shapes, and their contribution is proportional to different powers of $\{g^{(l)}(\tau)/g(\tau)^{l+1}\}$. This might seem trivial, because actually we have done an expansion in derivatives of $g(\tau)$. However, it is worth noting that because of the way we have set things up each coefficient $f_{i,j}(\wh)$ is a \emph{well-behaved} spectrum in $\wh$, without poles, and vanishing for $\wh \to \infty$ (no ultraviolet catastrophe).

\noindent

Finally, we have to say that, in general, we cannot guarantee the mathematical convergence of the adiabatic series, as we have formally integrated the series expansion of the Wightman function without any direct knowledge concerning its convergence. We can only be sure that this procedure leads to an asymptotic expansion in the dimensionless quantities $\{ g^{(l)}(\tau)/g(\tau)^{l+1} \}$.

\section{High-frequency behaviour}\label{high_energy}

All the terms $\Rs_n (\omega; g(\tau),g'(\tau),\ldots,g^{(n)}(\tau))$, and thus all the functions $f_{i,j} (\wh)$, tend to zero for $\omega \to \infty$, as can be  straightforwardly seen from their integral expressions~\eref{rate-coef}. Let us study in some detail the behaviour of their high-frequency falloff.
The integrals~(\ref{rate-coef}) that define the adiabatic expansion~(\ref{rate-series}) have the following asymptotic behaviour for $\omega \to \infty$ (see \cite{Satz:2006kb,wong_asymptotic_2001}):
\begin{multline}
\Rs_n \left(\omega, g(\tau),g'(\tau),\ldots,g^{(n)}(\tau) \right) \sim \\
2 \sum_{m=1}^\infty \left(-\fr{\omega^2}\right)^m W_n^{(2m-1)}\left(0, g(\tau),g'(\tau),\ldots,g^{(n)}(\tau) \right),
\label{asymptotic}
\end{multline}
where, using~\eref{Wn_terms_def}, the coefficients $W_n^{(l)}$ of this expansion can be written as
\begin{multline}
W_n^{(l)}\left(0, g(\tau),g'(\tau),\ldots,g^{(n)}(\tau) \right) = \\
\fr{n!}\evat{\pderiv[l]{}{s}\evat{\pderiv[n]{}{\alpha} \Wr \left(\tau+s,\tau,\alpha; g\right]}{\alpha \to 0^+}}{s=0}.
\label{derivs-coefs}
\end{multline}
Now, note from~\eref{new-Wreg} that the function $\Wr (\tau+s,\tau,\alpha;g]$ has the following scaling property:
\begin{equation}
Q^2\; \Wr \left(\tau+Qs,\tau,\alpha; g\right] = \Wr \left(\tau+s,\tau,Q\,\alpha; Q\, g \right], \quad Q \in \Rset.
\label{scaling}
\end{equation}
If we use this scaling property for $Q=-1$ and take into account  the invariance of the functional $\Wr$ with respect to the change $g \to -g$, we are led to the following result:
\begin{multline}
W_n^{(l)}\left(0; g(\tau),g'(\tau),\ldots,g^{(n)}(\tau) \right) = 0, \\
{\rm for} \quad n + l = 2p + 1,~p \in \Nset.
\label{odd-derivs}
\end{multline}
That is, any ``odd combination'' of derivatives with respect to $s$ and $\alpha$ will cancel when evaluated at $s = 0$ and $\alpha \to 0^+$. Noting that in~\eref{asymptotic} all the derivatives with respect to $s$ are odd, we conclude that all the terms in that asymptotic expansion cancel for $n$ even. That is, the even terms in the series expansion~\eref{rate-series} will decay to zero \emph{faster than any inverse power of $\omega$,} when $\omega \to \infty$. Also, the odd terms will contain only even inverse powers of $\omega$.

Also from the scaling property~\eref{scaling} we can show that the coefficients $W_n^{(l)}$ vanish for $l<n$. In order to obtain this result, let us first note that the order in which we take the derivatives with respect to $s$ and $\alpha$, and hence the evaluations at $\alpha=0$ and $s=0$,  in \eref{derivs-coefs} can be interchanged because we are considering real analytic functions $g(\tau)$, and thus $\Wr (\tau+s,\tau,\alpha; g]$ is also analytic. Second, because of \eref{scaling}, we can see that $\partial^l  \Wr/\partial s^l|_{s=0}$ is a polynomial in $\alpha$ of at most degree $l$. Indeed,
\begin{multline}
\pderivat[l]{\Wr \left(\tau+s,\tau,\alpha; g\right]}{s}{s=0} = \\
\evat{\pderivat[l]{\Wr \left(\tau+Qs,\tau,\alpha; g\right]}{s}{s=0}}{Q=1} = \\
\evat{\pderivat[l]{\Wr \left(\tau+Qs,\tau,\alpha; g\right]}{Q}{Q=0}}{s=1} = \\
\evat{\pderiv[l]{}{Q}\left[\fr{Q^2}\Wr \left(\tau+s,\tau,Q\alpha; Q g\right]\right]_{Q=0}}{s=1}.
\label{Q-derivs}
\end{multline}
(For dimensional consistency in this calculation, consider a quantity $T$ with dimensions of time, and then in the numerics set it to $T=1$.) Hence the result that $W_n^{(l)}$ vanish for $l<n$ follows directly noting that \eref{derivs-coefs} is proportional to the $n$th derivative of a polynomial of degree $l$ (at most).

This means that the expansion of the $n$th term in the adiabatic expansion $\Rs_n (\omega; g(\tau),g'(\tau),\ldots,g^{(n)}(\tau) )$ in inverse powers of $\omega$ will start from the term with $\omega^{-(n+1)}$. The adiabatic expansion up to order $n$ gives us automatically the high-energy behaviour up to order $n+1$. This relation is natural, because higher energies should depend crucially only on what happens in smaller intervals of time. And the slower the evolution of the acceleration is, the better a short interval of time ``represents'' what happens in larger intervals. This conceptually connects the adiabatic expansion with the high energy expansion.
In particular, up to fourth order the high energy behaviour is explicitly
\begin{align}
& \Rs_1 \left(\omega, g(\tau), g'(\tau) \right) \sim \frac{g(\tau) g'(\tau)}{24 \pi^2}\omega^{-2} \nonumber \\
& \hspace{4cm} + \frac{g(\tau)^3 g'(\tau)}{40 \pi^2}\omega^{-4} + O(\omega)^{-6}, \nonumber \\[10pt]
& \Rs_3 \left(\omega, g(\tau), g'(\tau), g''(\tau), g'''(\tau) \right) \sim \nonumber \\
& \hspace{0.1cm} - \fr{12 \pi^2}\left(\afifth g(\tau)g'''(\tau) + \ahalf g'(\tau)g''(\tau) \right)\omega^{-4} + O(\omega)^{-6}.
\label{high_dE}
\end{align}
Summing, we see
\begin{multline}
\Rs \left(\omega, \tau \right) \sim \frac{g(\tau) g'(\tau)}{24 \pi^2}\omega^{-2} + \fr{4 \pi^2}\left(\fr{10} g(\tau)^3 g'(\tau) \right. \\
\left. - \fr{15} g(\tau)g'''(\tau) - \asixth g'(\tau)g''(\tau) \right)\omega^{-4} + O(\omega)^{-6}.
\label{high_dE_total}
\end{multline}
The first term in the expansion fits with that found in~\cite{Louko:2006zv,Satz:2006kb}, while in~\cite{Obadia:2007qf} they already find terms up to order six, which also coincide with the terms in~\eref{high_dE_total}.

We have to say one final word about the high-energy expansion. In~\cite{Satz:2006kb} it is shown that, for a finite-time detector as described in subsection~\ref{real_detectors}, the response function falls off faster than any power of $\omega$ when $\omega \to \infty$. Thus, it seems that the high energy expansion we have found here is just an artifact of an infinite switching-on time, and a sharp switch-off. This is true if one considers the power expansion as the tendency of the response function \emph{for arbitrarily high energies.} As it is already mentioned in~\cite{Satz:2006kb}, the inverse power behaviour will always be cured with a much faster falloff in this limit, if the detector is switched carefully (see also~\cite{sriramkumar:1996}). But one can still use the high energy expansion just as a tool for calculating the response function \emph{for a fixed $\omega$} with arbitrarily good precision. Recalling considerations done in subsection~\ref{real_detectors}, as far as one makes sure that the effects due to the switching processes are negligible with respect to the precision desired, the use of the series is perfectly justified, and does give the response function for a finite-time detector.

\section{Numerical results}\label{example}

We will now consider a particular case for the acceleration function $g(\tau)$, given by
\begin{equation}
g(\tau) = \ahalf \left[ 1 + \tanh \left( \frac{\tau}{\dtau} \right) \right].
\label{g-sw}
\end{equation}
The detector begins with no acceleration in the asymptotic past, then starts to increase its acceleration smoothly and, after a transient period of about $\dtau$ in proper time, it asymptotically tends to reach a final acceleration that we will normalize to be equal to $1$ (which, together with $\hbar = c = 1$, determines a natural system of units for the problem). Note that this parameter $\dtau$ is closely related to the (inverse of the) parameter $\alpha$ used in order to obtain  the adiabatic expansion~\eref{rate-series}.

For this particular acceleration history the regularized Wightman function $W$ can be found analytically with the help of symbolic manipulation software (such as Mathematica). On the other hand, the response function needs to be calculated numerically, something we achieve with very high precision, thus providing a very good profile with which we can compare the various expressions obtained in this paper.

\subsection{Numerical response functions: slowly switching on acceleration}

\begin{figure}[ht]
\centering
\includegraphics[width=0.9\linewidth]{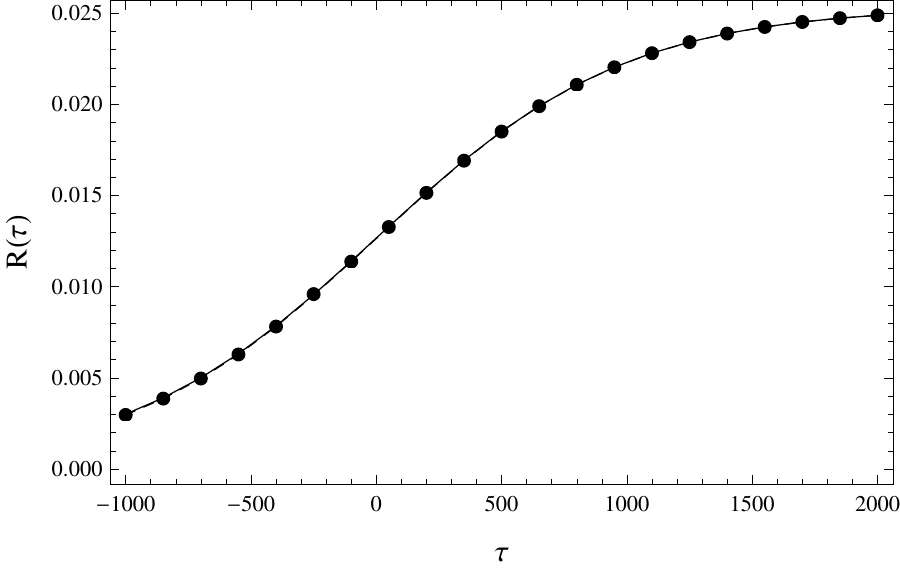}
\caption{Response function as a function of $\tau$ for $\omega = 0$, $\dtau = 1000$. Dots are the numerical solution, the solid line is the thermal spectrum approximation, and the dashed line (fully covered by the solid one) the third order adiabatic approximation.}
\label{dt-1000-En-0}
\end{figure}

\begin{figure}[ht]
\centering
\includegraphics[width=0.9\linewidth]{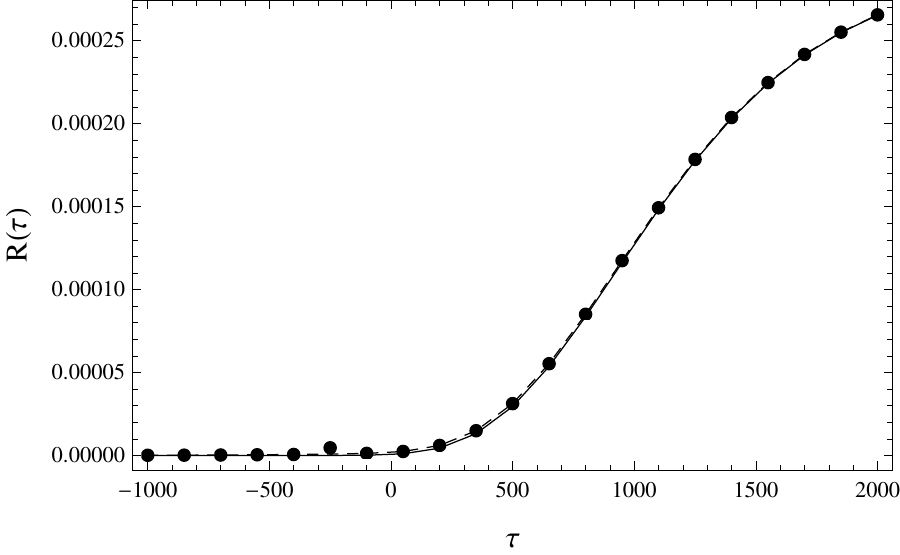}
\caption{Response function as a function of $\tau$ for $\omega = 1$, $\dtau = 1000$. Dots are the numerical solution, the solid line is the thermal spectrum approximation, and the dashed line (almost fully covered by the solid one) the third order adiabatic approximation.}
\label{dt-1000-En-1}
\end{figure}

\begin{figure}[ht]
\centering
\includegraphics[width=0.9\linewidth]{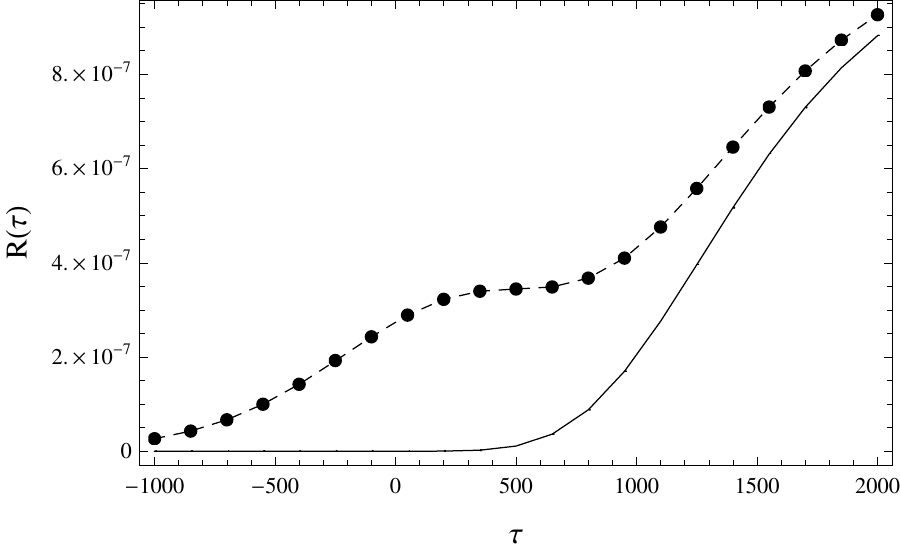}
\caption{Response function as a function of $\tau$ for $\omega = 2$, $\dtau = 1000$. Dots are the numerical solution, the solid line is the thermal spectrum approximation, and the dashed line the third order adiabatic approximation.}
\label{dt-1000-En-2}
\end{figure}

\begin{figure}[ht]
\centering
\includegraphics[width=0.9\linewidth]{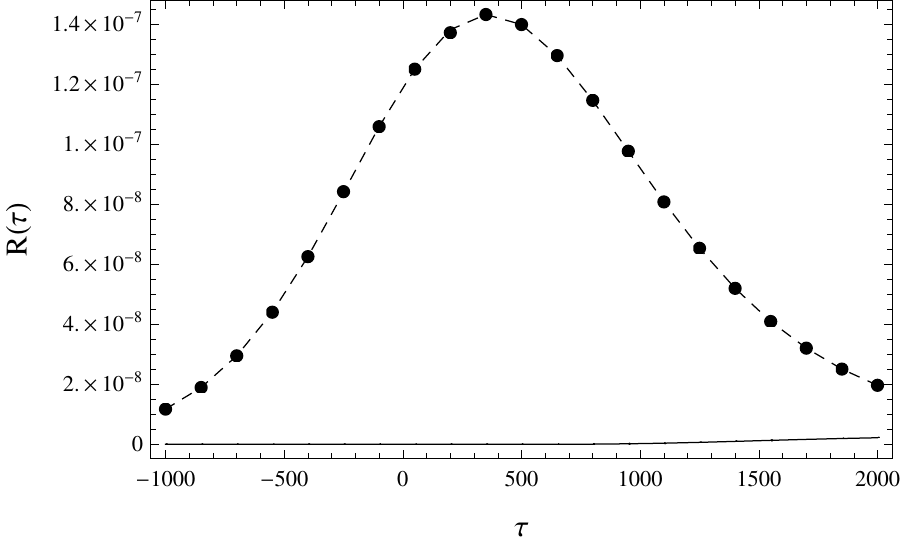}
\caption{Response function as a function of $\tau$ for $\omega = 3$, $\dtau = 1000$. Dots are the numerical solution, the solid line is the thermal spectrum approximation, and the dashed line the third order adiabatic approximation.}
\label{dt-1000-En-3}
\end{figure}

Let us describe some numerical results. First, consider the case $\dtau = 1000$, so that the acceleration is very slowly varying (in comparison to the timescale set by its asymptotic value, of the order of $1$). In figures~\ref{dt-1000-En-0} to~\ref{dt-1000-En-3}, we plot some graphs of the response function $\Rs(\omega,\tau)$ for different values of $\omega$. We also plot what would be the response function if the detector were just detecting a thermal bath with temperature proportional to $g(\tau)$, (the zeroth order approximation, as it is discussed in section~\ref{adiab}), and the result when using the adiabatic approximation up to the third order.

\noindent
\noindent
\noindent
\noindent

First of all, we see that, for all the energies considered, the fit with the adiabatic expansion up to the third order is virtually perfect. On the other hand, for low energies the match with the thermal detection seems to be perfect, while for higher energies the results start to differ. This might seem surprising when taking into account the results found in subsection~\ref{high_energy}: The adiabatic expansion up to order $n$ coincides with the high energy expansion up to order $n+1$, so that the higher the energies that we measure, the better the thermal approximation will be. In particular, the zeroth order adiabatic approximation \emph{must be exact} for arbitrarily high energies. And this is true, simply because both the exact result and the approximation are actually zero. But it is true \emph{for the absolute difference between the exact value and the approximation.} What figures~\ref{dt-1000-En-0} to~\ref{dt-1000-En-3} make evident is the \emph{relative difference,} in comparison with the magnitude of the exact result (see the values in the vertical axis). And although the absolute difference between the numerical results and the thermal curve for arbitrarily low energies (figure~\ref{dt-1000-En-0}) \emph{is greater} than for energies $\omega \sim 3$ (figure~\ref{dt-1000-En-3}), in a relative sense the difference is negligible for low energies, while for high energies it is much more important. Moreover, for energies still of the order of $\omega \sim 3$, it is the contribution of the zeroth order term to the total spectrum which is actually negligible with respect to higher order contributions. In particular, the ``bell shape'' that appears in that figure corresponds, to a very good approximation, to the first term in the high energy expansion~\eref{high_dE_total}, proportional to $g(\tau)g'(\tau)$.

Finally, note also that, for $\tau \to \pm \infty$, the three curves plotted in all the figures converge to each other. This is to be expected, since in this region the acceleration approaches a constant value, and thus the zeroth order adiabatic approximation must tend to the exact result.

\subsection{Numerical response functions: rapidly switching on acceleration}
\begin{figure}[ht]
\centering
\includegraphics[width=0.9\linewidth]{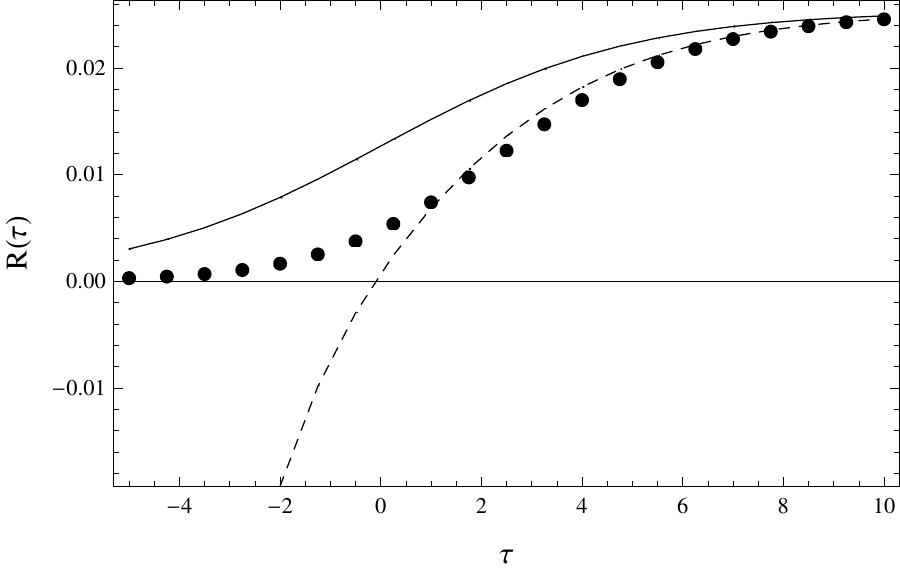}
\caption{Response function as a function of $\tau$ for $\omega = 0$, $\dtau = 5$. Dots are the numerical solution, the solid line is the thermal spectrum approximation, and the dashed line the third order adiabatic approximation.}
\label{dt-5-En-0}
\end{figure}

\begin{figure}[ht]
\centering
\includegraphics[width=0.9\linewidth]{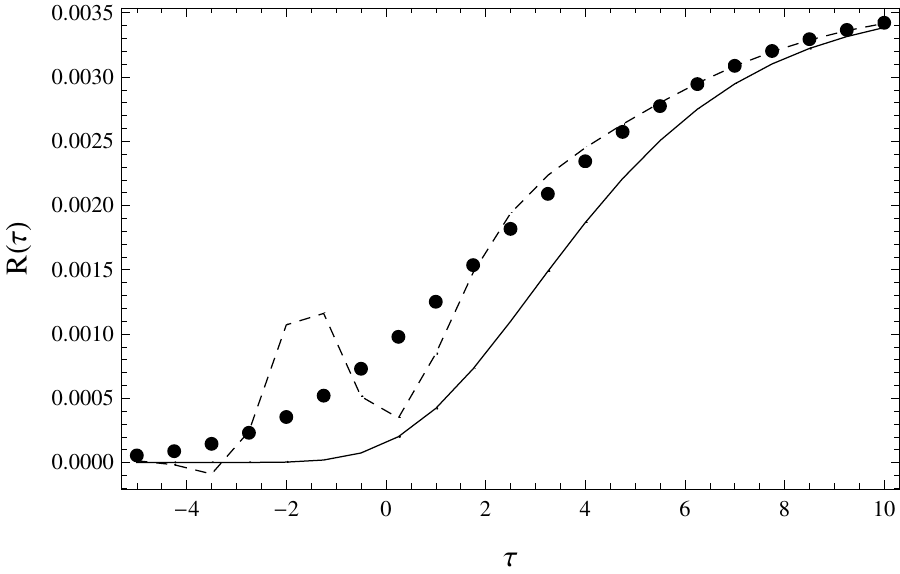}
\caption{Response function as a function of $\tau$ for $\omega = 0.5$, $\dtau = 5$. Dots are the numerical solution, the solid line is the thermal spectrum approximation, and the dashed line the third order adiabatic approximation.}
\label{dt-5-En-05}
\end{figure}

\begin{figure}[ht]
\centering
\includegraphics[width=0.9\linewidth]{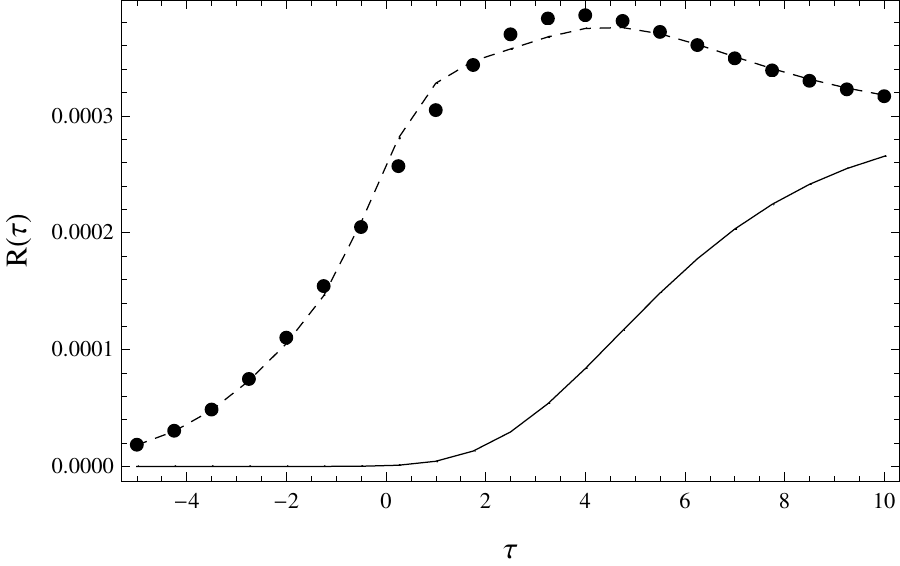}
\caption{Response function as a function of $\tau$ for $\omega = 1$, $\dtau = 5$. Dots are the numerical solution, the solid line is the thermal spectrum approximation, and the dashed line the third order adiabatic approximation.}
\label{dt-5-En-1}
\end{figure}

\begin{figure}[ht]
\centering
\includegraphics[width=0.9\linewidth]{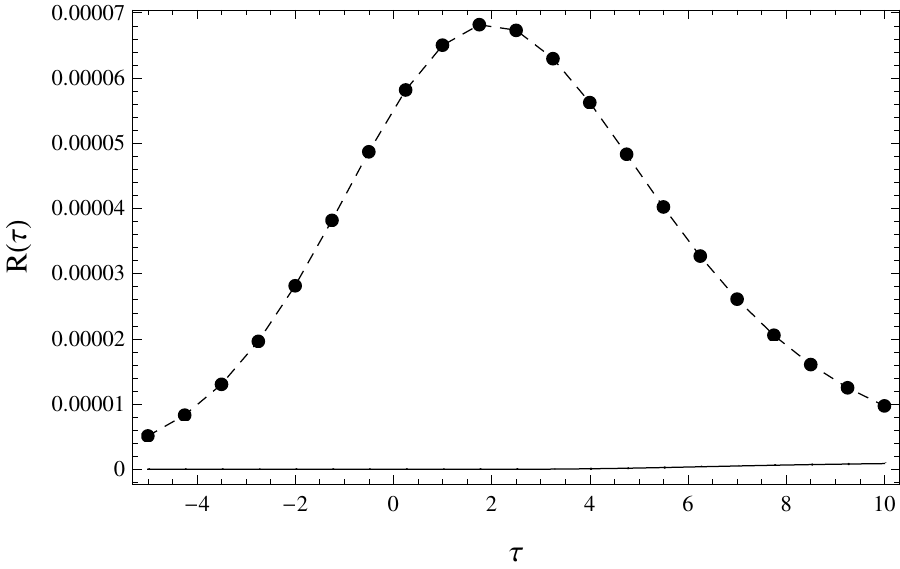}
\caption{Response function as a function of $\tau$ for $\omega = 2$, $\dtau = 5$. Dots are the numerical solution, the solid line is the thermal spectrum approximation, and the dashed line the third order adiabatic approximation.}
\label{dt-5-En-2}
\end{figure}

We will now plot the response function for the case $\dtau = 5$ (figures~\ref{dt-5-En-0} to~\ref{dt-5-En-2}).

\noindent
\noindent
\noindent
\noindent

In this case, we are much further away from the adiabatic regime. In contrast to the previous case, even for arbitrarily low energies, deviations from the thermal response function can be clearly seen. The adiabatic approximation (up to third order) starts to fail for lower energies, but it is still virtually exact for higher energies. Here differences with the third order approximation are bigger for lower energies \emph{even in a relative sense.} This is again in agreement with the fact that adiabatic expansion gives ``for the same price'' the high energy expansion. Note in particular that in figure~\ref{dt-5-En-0} the third order expansion seems to be a worse approximation than the zeroth order. This is what one must expect when staying far from the adiabatic regime, as we are dealing just with an asymptotic series,  without (in general) any guarantee of convergence.

\section{Summary}\label{summary}

In this article, we have first calculated the transition probability for an Unruh--DeWitt detector in Minkowski spacetime, when interacting with a Klein--Gordon massless real scalar field in its Minkowski vacuum state. The detector is set to be in its ground state in the asymptotic past, and we calculate the transition probability when it follows a trajectory $x(\tau')$ up to some time $\tau$. We then consider a cumulative multipart macro-detector, as an ensemble of many Unruh--DeWitt micro-detectors with suitably small coupling constant. This allows us to use first-order perturbation theory to calculate the average number of detectors that are excited at some time $\tau$. Ignoring the detector's internal details, and taking the derivative with respect to the time $\tau$, we calculate what we have called the response function $\Rs (\omega, \tau)$ at some time $\tau$ and some energy $\omega$ of the detected particle. We also justify why, and up to what precision, we can consider an idealized situation with an infinite switching time (from the remote past) and a sharp switch-off of the detector at $\tau$ when calculating the response function.

Using the general expression for accelerated trajectories in one dimension, we manage to write the response function as a functional of the acceleration $g(\tau')$. Then, we expand in powers of a formal adiabaticity parameter, which is inserted \emph{by hand} to control how fast the acceleration changes. After calculating the expansion in terms of this formal parameter, we remove it from each term in the expansion. The zeroth order term is the thermal spectrum with temperature proportional to the instantaneous acceleration. That is the well-known Unruh effect result. Higher-order terms involve increasingly higher-order derivatives of the acceleration function, (jerk, snap, etcetera).  

We see that all the individual terms in the adiabatic expansion tend to zero in the high-frequency limit. There is no ultraviolet catastrophe for any of these terms. We also see that the even-order terms decay to zero much faster than any inverse power of $\omega$, while odd terms of order $n$ in the adiabatic expansion are of order $\omega^{-(n+1)}$. In other words, the adiabatic expansion to order $n$ automatically gives  the high-frequency expansion to order $n+1$.

Finally, we use the adiabatic expansion with a simple but representative example of switch-on function for the acceleration. The numerical results obtained reflect most of the aspects we have discussed in the article. This example shows that the adiabatic expansion is a useful tool for generalizing the study of the Unruh effect to trajectories with variable acceleration. Further generalizations could include (among many other possibilities) more than one-dimensional trajectories, non-monopole coupling with the field to study possible anisotropies in the radiation, or the same problem for a field theory in a curved spacetime.

\section{Acknowledgements}

The authors wish to thank Carlos Barcel\'o and Luis J. Garay for enlightening discussions. Financial support (LCB) was provided by the Spanish MINECO through the projects  FIS2011-30145-C03-01 and by the Junta de Andaluc\'{\i}a through the project FQM219. MV was supported by a Marsden Grant and a James Cook Fellowship, both administered by the Royal Society of New Zealand. 


\end{document}